\documentclass[12pt]{article}
\pdfoutput=1
\usepackage{amssymb,amsmath}
\usepackage{amsbsy}   
\usepackage{amsfonts}
\usepackage{graphicx}
\usepackage{cancel}
\usepackage{enumerate}
\usepackage{collref}
\collectsep{;\ }
\usepackage[colorlinks=true]{hyperref}
\newcommand{\kd}{\kappa^{D}}
\newcommand{\rd}{\rho^{D}}
\newcommand{\ku}{\kappa^{U}}
\newcommand{\ru}{\rho^{U}}
\newcommand{\kl}{\kappa^{L}}
\newcommand{\rl}{\rho^{L}}

\newcommand{\sba}{s_{\beta-\alpha}}
\newcommand{\cba}{c_{\beta-\alpha}}

\newcommand{\AddrUdeA}{
  {\it Instituto de F\'{\i}sica, Universidad de Antioquia, Medell\'\i n, Colombia}}
\newcommand{\AddrUNAL}{
  {\it Departamento de F\'{\i}sica, Universidad Nacional de Colombia, Bogot\'a, Colombia}}

\title{The role of charged Higgs boson decays in the determination \\ of $\tan\beta$-like parameters}

\author{H. Cardenas$^1$, D. Restrepo$^2$, J.-Alexis Rodriguez$^1$\\
1.\AddrUNAL \\ 2.\AddrUdeA}
\date{}
\begin{document}
\maketitle{}

\begin{abstract}

The most commonly used parameterizations of the Yukawa couplings in the two Higgs doublet model are revisited. Their similitudes
and differences are emphasized and relations between the different notations used in the literature are derived. Using these relations, bounds on the space of parameters
of the charged Higgs sector of the two Higgs doublet model are obtained. These constraints are obtained from flavor observables like the measurement of
 $B \to X_s \gamma$ and the recent measurement of $B_s \to \mu^+ \mu^-$ from LHCb. The ratio 
 $R_ {H^+}={B(H^+ \to \tau^+ \bar \nu_\tau)}/{B(H^+ \to t \bar b)}$ is evaluated
for the two Higgs doublet model type-III and type-II, and their differences are quantified.
\end{abstract}

One simple framework beyond the standard model (SM) consists in adding a Higgs doublet, leading to a rich Higgs sector
spectrum and therefore, giving a richer phenomenology than in the SM \cite{Lee:1973iz,Glashow:1976nt}. This model is usually called the Two Higgs doublet
Model (THDM). In the THDM there are three Goldstone bosons that give masses to the gauge bosons and there are also
five Higgs bosons:  two CP-even neutral Higgs bosons, one CP-odd neutral Higgs boson and two charged Higgs bosons. 

There are multiple ways to implement this kind of SM extension. These ways depend on how
 the couplings between the Higgs bosons and the fermions are chosen. They should be chosen carefully in order to avoid  Flavor 
Changing Neutral Currents (FCNC) which are generated at the tree level in the most general version of the THDM, the so 
called type-III THDM. In contrast, models like the type-I and II THDM impose an additional discrete symmetry in order to avoid
the FCNC at tree level \cite{Glashow:1976nt}. In the type-I model, the couplings between the Higgs bosons and the fermions  are chosen such that only one Higgs doublet 
couples to the SM fermions, whereas in the type-II model, one doublet couples with the up quark sector and the another
 one with the down quark sector. In the type-III model, although both Higgs doublets couple with
 the SM fermions, the model is still phenomenologically possible if the FCNC at tree level are suppressed \cite{Atwood:1996vj}.

The THDM type-II has been studied  more extensively in the literature, motivated by its similarities in the Higgs sector at tree level with the Minimal Supersymmetric Standard Model (MSSM), which requires a second Higgs doublet.  However, due to quantum loop corrections, the Higgs sector will actually be described by an effective potential
which is similar to the most general version of the THDM, the so-called type-III.

From a phenomenological point of view, it is important to study the properties of the most general THDM. In this  
model, there are two Yukawa matrices per fermion type (up or down type quarks and leptons) which cannot be diagonalized
simultaneously. There are two different ways to deal with the FCNC couplings: the first one is to adopt a particular 
Yukawa texture like the Cheng-Sher ansatz \cite{Cheng:1987rs} which assumes that the couplings are proportional to the geometric mean
of the  two masses ($\xi_{ij} \sim \sqrt{m_i m_j}$); the second one is to impose the Minimal Flavor Violation (MFV) hypothesis \cite{D'Ambrosio:2002ex}
 as it was already done in the so-called Aligned Two Higgs Doublet Model (ATHDM) \cite{Pich:2009sp,Jung:2010ik}, where the two Yukawa matrices per fermion 
are proportional to each other and then, they can be diagonalized avoiding the FCNCs at tree level. The alignment condition holds at high
energy, but  it has been shown that it does not hold at low energies because of the quantum corrections concerning
 the quark sector \cite{Jung:2010ik}.

Different parameterizations of the couplings  have been suggested. Haber and Davidson \cite{Davidson:2005cw} propose three
different $\tan \beta$-like parameters,  associated to the up, down and charged lepton sector. They also emphasize that,
the parameter $\tan \beta=v_2/v_1$ of the MSSM is not a physical parameter of the THDM type-III, because it corresponds
to an specific basis of the Higgs doublets. In the framework of the ATHDM \cite{Pich:2009sp,Jung:2010ik}, there are also three 
parameters 
$\zeta_f$ related to the up, down and lepton sector. In addition, there is also the version by Ibarra, {\it et.al.} \cite{Braeuninger:2010td} 
who defined three
different angles relative to the parameter $\tan \beta=v_2/v_1$ of the THDM type-II. In all cases, efforts
 have been done 
to systematically restrict the available space of parameters of the model under consideration using the known phenomenology.
Our aim is to compare these different parameterizations and to try to unify all these studies in terms of 
the most convenient parameters. These  parameters  should be
such that they can be easily compared to the THDM type-II in order to take it as reference and differentiate between the models.

The THDM is an $SU(3)_C \otimes SU(2)_L\otimes U(1)_Y$ gauge model with the particle spectrum of the standard model plus two scalar doublets with hypercharge
$Y= \frac 12$. In general, both neutral components of the two scalar fields can acquire vacuum expectation values (VEV) different from zero. However, using a global $SU(2)$
transformation, it is possible to choose that only one of them has non-zero VEV. Then, the three Goldstone fields are in just one doublet,
\begin{equation}
 H_1=\begin{pmatrix}
    G^+\\
    \frac{1}{\sqrt{2}}(v+S_1+iG_0)\\
  \end{pmatrix} , \,\, 
H_2=\begin{pmatrix}
    H^+\\
    \frac{1}{\sqrt{2}}(S_2+iS_3)\\
  \end{pmatrix}.
\end{equation}
In this form, the scalar particle spectrum consists of the charged fields $H^\pm$ and the neutral fields $\phi_0=(h,H,A)$ which in general can be written as  $\phi_{0i}=R_{ij}S_j$, where the orthogonal
matrix $R_{ij}$ is defined by the scalar potential. If CP invariance is assumed then $A=S_3$ and
\begin{equation}
\label{defmix}
 \begin{pmatrix}
    H\\
    h\\
  \end{pmatrix}=
  \begin{pmatrix}
    \cos(\alpha-\beta)&\sin(\alpha-\beta)\\
    -\sin(\alpha-\beta)&\cos(\alpha-\beta)\\     
  \end{pmatrix}
  \begin{pmatrix}
    S_1\\
    S_2    

  \end{pmatrix}\, \, ,
\end{equation}
where the angle $(\alpha-\beta)$ appears as a physical parameter.

The most general expression for the Lagrangian in the THDM is
\begin{align}
\label{eq:1}
-{\cal L}_{Y}=&\overline{Q^0}_{L}\eta^{U,0}_1\tilde{\Phi}_{1}U^0_{R}+\overline{Q^0}_{L}\eta^{D,0}_1\Phi_{1}D^0_{jR}+
\overline{Q^0}_{L}\eta^{U,0}_2\tilde{\Phi}_{2}U^0_{R}+\overline{Q^0}_{L}\eta^{D,0}_2\Phi_{2}D^0_{R}\nonumber\\
& 
+\overline{L^0}_{L}\eta^{E,0}_1\Phi_{1}E^0_{R}+\overline{L^0}_{L}\eta^{E,0}_2\Phi_{2}E^0_{R}+ \text{h.c.,}
\end{align}
where $\eta^{U,D}_{i}$ are non-diagonal mixing matrices $3\times3$, $\tilde\Phi_i=i\sigma_2\Phi_i$, $(U,D)_R$ are 
right-handed fermion singlets, $Q_L$ are left-handed fermion doublets, and the index 0 indicates that the fields are not mass eigenstates.
The Lagrangian, in the Higgs basis and using the definitions 
\begin{align}
  \label{eq:2}
  \kappa^{U,0}=& \eta^{U,0}_1c_\beta+\eta^{U,0}_2e^{-i\theta}s_\beta \, \, ,\nonumber\\
  \rho^{U,0}=&-\eta^{U,0}_1s_\beta+\eta^{U,0}_2c_\beta e^{-i\theta} \, \, , \nonumber\\
  \kappa^{F,0}=&\eta_1^{F,0}\cos\beta+\eta_2^{F,0}e^{i\theta}\sin\beta \, \, , \nonumber\\
  \rho^{F,0}=&-\eta_1^{F,0}\sin\beta+\eta_2^{F,0}e^{i\theta}\cos\beta\, ,
\end{align}
with ($F=D,E$), can be written as
\begin{align}
  \label{eq:7}
 -\mathcal{L}_Y =&\overline{Q^0}_{L}\kappa^{U,0}\tilde H_1U^0_{R}+\overline{Q^0}_{L}\kappa^{D,0}H_1D^0_{R}
+\overline{Q^0}_{L}\rho^{U,0}\tilde H_2U^0_{R}+\overline{Q^0}_{L}\rho^{D,0}H_2D^0_{R}\nonumber\\
&+\overline{L^0}_L H_1\kappa^{E,0}E^0_R+\overline{L}_L H_2\rho^{E,0}E^0_R +\text{h.c.}
\end{align}

Generally, the mass matrices in Eq.~(\ref{eq:1}) cannot be diagonalized simultaneously. We define
the mass basis in the usual way as
\begin{align}
  U_L=& V_L^UU_L^0 & U_R=V_R^UU_R^0 \nonumber \\
  D_L=& V_L^DD_L^0 & D_R=V_R^DD_R^0 \nonumber \\
  E_L=& V_L^EE_L^0 & E_R=V_R^EE_R^0\,.
\end{align}
For example, the second term of the Lagrangian turns into 
\begin{align*}
  -\mathcal{L}_Y\supset&\overline{Q}_L^0 H_1\kappa^{D,0}D^0_R\\
=&\overline{U}_L^0H_1^1\kappa^{D,0}D^0_R+\overline{D}_LH_1^2\kappa^{D,0}D^0_R\\
=&\overline{U}_LH_1^1V_L^U\kappa^{D,0}{V_R^D}^\dagger D_R+\overline{D}_LH_1^2V_L^D\kappa^{D,0}{V_R^D}^\dagger D_R\\
=&\overline{U}_LH_1^1V_L^U{V_L^D}^\dagger V_L^D\kappa^{D,0}{V_R^D}^\dagger D_R+\overline{D}_LH_1^2\kappa^{D}D_R\\
=&\overline{U}_LH_1^1K\kappa^{D}D_R+\overline{D}_LH_1^2\kappa^{D}D_R
\end{align*}
where  $K=V_L^U{V_L^D}^\dagger$  and  $\kappa^{D}=V_L^D\kappa^{D,0}{V_R^D}^\dagger$ is the corresponding diagonal matrix.  
In an arbitrary basis, the Yukawa matrices for the Higgs doublets  are given by~\cite{Davidson:2005cw}
\begin{equation}
  \begin{aligned}
    -\mathcal{L}_{\rm{Y}}&=\frac{1}{\sqrt{2}}\overline{D}\Bigl[\kd\sba+\rd\cba \Bigr]Dh
    +\frac{1}{\sqrt{2}}\overline{D}\Bigl[\kd\cba-\rd\sba \Bigr]DH+ \frac{\mathrm{i}}{\sqrt{2}}\overline{D}\gamma_5\rd DA \\
    &\,+\frac{1}{\sqrt{2}}\overline{U}\Bigl[\ku\sba+\ru\cba \Bigr]Uh
    +\frac{1}{\sqrt{2}}\overline{U}\Bigl[\ku\cba-\ru\sba \Bigr]UH- \frac{\mathrm{i}}{\sqrt{2}}\overline{U}\gamma_5\ru UA \\
    &\,+\frac{1}{\sqrt{2}}\overline{L}\Bigl[\kl\sba+\rl\cba \Bigr]Lh
    +\frac{1}{\sqrt{2}}\overline{L}\Bigl[\kl\cba-\rl\sba \Bigr]LH+ \frac{\mathrm{i}}{\sqrt{2}}\overline{L}\gamma_5\rl LA \\
    &\,+\frac{}{}\Bigl[\overline{U}\bigl(V_{\rm{CKM}} \rd P_R-\ru V_{\rm{CKM}} P_L\bigr)DH^+ + \overline{\nu}\rl P_RL H^+ + \rm{h.c.}\Bigr].                                                                                              
\end{aligned}
\label{eq:yukawa}
\end{equation}
It is worth noting that eq.~(\ref{eq:yukawa}) exhibits tree-level FCNCs unless the $\rho^F$ are diagonal. A sufficient condition is that each fermion type $F=\{D,U,L\}$ 
couples to only one Higgs doublet \cite{Glashow:1976nt}, this is equivalent to $\eta^{F,0}_1=0$ or $\eta^{F,0}_2=0$, which 
leads to the relations $\rho^{F,0}=\kappa^{F,0}\cot\beta$ and $\rho^{F,0}=-\kappa^{F,0}\tan\beta$, respectively.

For this Lagrangian, Haber and Davidson \cite{Davidson:2005cw} have introduced three $\tan\beta$--like parameters $\tan\beta_{D,E}$ and $\cot\beta_U$
 in the case of one quark/lepton generation.
The construction of $\tan\beta$--like parameters in the multi-generation THDM-III is more complicated \cite{Davidson:2005cw}. 
Fortunately, in the phenomenological viable models, the third generation Yukawa couplings dominate, and one may define 
$\tan\beta$--like parameters based only on the consideration of third generation fermion couplings. One 
alternative, is just to consider the $\tan\beta$ like parameters as the 
three alignment parameters to suppress FCNCs.

In the Cheng \& Sher \cite{Cheng:1987rs} parameterization where the couplings are proportional to the geometric mean
of the  two masses $\xi_{ij} =\lambda_{ij} \sqrt{m_i m_j}/v$, the third generation Yukawa couplings  dominate,
 and it is possible to define $\tan\beta$--like parameters from the relations
\begin{align}
\label{chengsher}
  \rho^f_{33}=&\lambda^f_{33}\frac{m_{33}}{v}=\lambda^f_{33}\kappa^f_{33}\,\, ,\nonumber\\
 Y_f=&\lambda_{33}^f M_f\,,
\end{align}
where $Y_f=v\rho^f_{33}$ and $M_f=m_{33}^f$\footnote{Cheng \& Sher \cite{Cheng:1987rs} parameterization can be written in general in terms of nine $\tan\beta$--like parameters~\cite{Mahmoudi:2009zx}.}.
Therefore, it is possible to establish a relationship between the Haber and Davidson \cite{Davidson:2005cw} parameterization and the one
by Cheng and Sher \cite{Cheng:1987rs}.

On the other hand, the alignment hypothesis  considers \cite{Pich:2009sp,Jung:2010ik}
\begin{align}
  \label{eq:3}
 \eta_2^{U,0}=&\xi_U^*e^{i\theta}\eta_1^{U,0} & \eta_2^{F,0}=&\xi_fe^{-i\theta}\eta_1^{F,0}\,,
\end{align}
where $F=D,E$. Therefore, using the expressions (\ref{eq:2}) and defining 
\begin{align}
\label{eq:5}
  M_f'=&\frac{v}{\sqrt{2}}\kappa^{f,0} & Y_f'=&\frac{v}{\sqrt{2}}\rho^{f,0}\,, 
\end{align}
where $f=U,D,E$, the following relations can be obtained:
\begin{align}
  \label{eq:12}
   M_F'=&\frac{1}{\sqrt{2}}v_1(1+\xi_F\tan\beta)\eta_1^{F,0}\,, \nonumber \\
  Y'_F=&\frac{1}{\sqrt{2}}v_1(-\tan\beta+\xi_F)\eta_1^{F,0}\nonumber\\
    =&M_F'\frac{\xi_F-\tan\beta}{1+\xi_F\tan\beta}\nonumber\\
    =&\zeta_FM_F'\,.
\end{align}
Similarly for the up-sector
\begin{align}
  Y_U'=\zeta_U^*M_U'\,,
\end{align}
where ($f=U,D,E)$ and
\begin{align}
  \label{eq:10}
  \zeta_f=\frac{\xi_f-\tan\beta}{1+\xi_f\tan\beta} \, \, .
\end{align}
On that ground, if the alignment condition is fulfilled, the $Y_f'$ and $M_f'$ are proportional to each other\cite{Pich:2009sp,Jung:2010ik} and therefore, they can 
be simultaneously diagonalized. The next step then is to write down the Yukawa Lagrangian in the mass basis, which is
given by $M_f=V_L^f M_f' V_R^f$, $Y_{D,E}=\zeta_{D,E}M_{D,E}$ and  $Y_U=\zeta^*_U M_U$, then
\begin{align}
   -\mathcal{L}_Y=&\frac{\sqrt{2}}{v}\left(
      \overline{Q}_L\tilde H_1M_UU_R+\overline{Q}_LH_1M_DD_R
      +\zeta_U^*\overline{Q}_L\tilde H_2Y_UU_R
      +\zeta_D\overline{Q}_LH_2Y_D D_R\right.\nonumber\\
    &+\left.
      \overline{L}_LH_1M_EE_R+\zeta_E\overline{L}_LH_2Y_EE_R
  +\text{h.c}\right).
\end{align}

At this point, it is useful to analyze separately the charged Higgs and the
neutral Higgs interactions. The charged Higgs interactions take the form
\begin{align}
  \mathcal{L}_{H^+}=-\frac{\sqrt{2}}{v}\overline{U}(\zeta_D K M_D P_R-\zeta_U M_U K P_L)D H^++\text{h.c}\,.
\end{align}
while the neutral Higgs interaction is
\begin{align}
  &\mathcal{L}_{h\overline{D}D}=-\frac{1}{v}\overline{D}y_D^hM_DP_RD+\text{h.c}\nonumber\\
&=-\frac{1}{v}\overline{D}M_D[R_{11}+(R_{12}+iR_{13})\zeta_D]P_RD+\text{h.c}\nonumber\\
&=-\frac{1}{v}\overline{D}M_D[R_{11}+(R_{12}+iR_{13})\zeta_D]P_RD -\frac{1}{v}(P_RD)^\dagger M_D^*[R_{11}^*+(R_{12}^*-iR_{13}^*)\zeta_D^*]\gamma^0D\nonumber\\
&=-\frac{1}{v}\overline{D}M_D[R_{11}+(R_{12}+iR_{13})\zeta_D]P_RD -\frac{1}{v}D^\dagger P_R M_D[R_{11}^*+(R_{12}^*-iR_{13}^*)\zeta_D^*]\gamma^0D\nonumber\\
&=-\frac{1}{v}\overline{D}M_D[R_{11}+(R_{12}+iR_{13})\zeta_D]P_RD -\frac{1}{v}\overline{D} M_D[R_{11}^*+(R_{12}^*-iR_{13}^*)\zeta_D^*]P_LD\,.
\end{align}
Assuming a CP-conserving Higgs potential, or equivalently the absence of tree--level mixing
 between the CP-odd Higgs bosob ($A$) and the CP-even Higgs bosons ($h$ , $H$) \cite{Davidson:2005cw}, the $R_{ij}$ elements are real and 
$R_{13}=0$ (see equation~(\ref{defmix})). Therefore, the Lagrangian can be reduced to
\begin{align}
  \mathcal{L}_{h\overline{D}D}=&-\frac{1}{v}\overline{D}M_D[R_{11}+R_{12}\zeta_D]P_RD
-\frac{1}{v}\overline{D} M_D[R_{11}+R_{12}\zeta_D^*]P_LD\nonumber\\
 =&-\frac{1}{v}\overline{D}M_D[s_{\beta-\alpha}+\zeta_Dc_{\beta-\alpha}]D\,.
\end{align}
using $R_{11}=s_{\beta-a}$, $R_{12}=c_{\beta-\alpha}$ and taking $\zeta_D$ as real.

It is worth to take into account the following relationship
\begin{align}
  \lim_{\zeta_D\to-\tan\beta}(s_{\beta-\alpha}+\zeta_Dc_{\beta-\alpha})=-\frac{\sin\alpha}{\cos\beta} \, ,
\end{align}
and comparing to the notation by Haber and Davidson \cite{Davidson:2005cw}, it can be seen that the parameters $\zeta_{U,D,E}$ of the alignment
 model \cite{Pich:2009sp,Jung:2010ik} are
equivalent to the parameters $\tan \beta_{U,D,L}$ since
\begin{align}
  \tan\beta_D=&-\zeta_D \, ,\nonumber\\
  \tan\beta_L=&-\zeta_E \, ,\nonumber\\
  \cot\beta_U=&\zeta_U \, \, .
\end{align}

There is another parameterization of the alignment condition given by Ibarra,{\it et.al} \cite{Braeuninger:2010td}, in terms of phases
relative to the usual parameter $\tan \beta$. In order to do the comparison, the Yukawa Lagrangian
from Eq.~\eqref{eq:1} can be re-written taking into account the relations
\begin{align}
 M'_f=&\frac{1}{\sqrt{2}}\eta_1^{f,0}v_1+\frac{1}{\sqrt{2}}\eta_2^{f,0}v_2\nonumber \,\\
 Y'_f=&-\frac{1}{\sqrt{2}}\eta_1^{f,0}v_2+\frac{1}{\sqrt{2}}\eta_2^{f,0}v_1\,.
\end{align}
Thus, this parameterization is interpreted at the high energy cut--off scale, $\Lambda$, as:
\begin{align}
  \label{eq:15}
  \eta_1^{f,0}(\Lambda)=&\cos\psi_f Y''_f & \eta_2^{f,0}(\Lambda)=&\sin\psi_f Y''_f \, .
\end{align}
The above equations for the not-well 
defined Yukawa $Y_f''$ should be solved to recover the alignment definition. Hence, 
the alignment hypothesis assuming real parameters is
\begin{align}
  \eta_2^{f,0}=\tan\psi_f \eta_1^{f,0}\,.
\end{align}
After comparing with Eq.~\eqref{eq:3}, it is obtained $\tan\psi_f=\xi_f$.

The mass and Yukawa matrices in the Higgs basis, Eqs. \eqref{eq:12}, can also be 
rewritten as
\begin{align}
   M_f'=&\frac{1}{\sqrt{2}}v_1(1+\tan\psi_f\tan\beta)\eta_1^{f,0}\,,
\end{align}
\begin{align}
\label{eq:16}
  Y'_f=&\frac{1}{\sqrt{2}}v_1(-\tan\beta+\tan\psi_f)\eta_1^{f,0}\nonumber\\
    =&M_f'\frac{\tan\psi_f-\tan\beta}{1+\tan\psi_f\tan\beta}\nonumber\\
    =&\tan(\psi_f-\beta)M_f'\,,
\end{align}
which are later compared with Eq.~\eqref{eq:9} and the following relation is obtained
\begin{align}
  \label{eq:17}
  \tan(\beta-\psi_f)=-\zeta_f \, .
\end{align}
On the other hand,
\begin{align}
  \zeta_f=\tan(\psi_f-\beta)=\frac{\tan\psi_f-\tan\beta}{1+\tan\psi_f\tan\beta}
\end{align}
which can be compared with Eq.~\eqref{eq:10}, and as expected, it is 
\begin{align}
  \tan\psi_f=\xi_f\,.
\end{align}

Considering that the alignment condition is satisfied ($Y_f'$ and $M_f'$ are proportional) 
and  using $M_f=V_L^f M_f' V_R^f$ then
\begin{align}
\label{eq:9}
  Y_f=&-\tan(\beta-\psi_f)M_f\,.
\end{align}
Notice that  now it is straight formward to recover the THDM-II:  $\psi_u=\pi/2$ implies
\begin{align}
  \tan(\pi/2-\beta)=\lim_{\tan\psi_u \to\infty}\zeta_U=\frac{1}{\tan\beta}=\cot\beta\,,
\end{align}
moreover $\psi_{d,l}=0$ suggests
\begin{align}
  \zeta_{F}=-\tan\beta\, ;
\end{align}
instead of the established relationship in the original paper \cite{Braeuninger:2010td}. Finally, the Yukawa Lagrangian
in terms of the mass eigenstates is
\begin{align}
  & -\mathcal{L}_Y=\frac{\sqrt{2}}{v}\left(
      \overline{Q}_L\tilde H_1M_UU_R+\overline{Q}_LH_1M_DD_R
      +\tan(\psi_u-\beta)\overline{Q}_L\tilde H_2Y_UU_R \right.\nonumber\\
    &+\left.  \tan(\psi_d-\beta)\overline{Q}_LH_2Y_D D_R 
     \overline{L}_LH_1M_EE_R+\tan(\psi_e-\beta)\overline{L}_LH_2Y_EE_R
  +\text{h.c}\right).
\end{align}

The parameterization by Ibarra {\it et.al} \cite{Braeuninger:2010td} is interesting because it allows to recover a THDM type-II and preserves
the parameter $\tan \beta$, thus convenient for phenomenological analyses. Moreover, in order to keep this recovery mode easier
than in the previous definitions, it is convenient to re-define the alignment conditions for the up sector as
\begin{align}
  \eta_2^{U,0}(\Lambda)=&-\cot \psi_U \eta_1^{U,0}(\Lambda) \,,
\end{align}
which produces
\begin{align}
  M'_U=&\frac{v}{\sqrt{2}}\left(\eta_1^{U,0}\cos\beta+\eta_2^{U,0}\sin\beta\right)\nonumber\\
  =&\frac{v}{\sqrt{2}}\eta_1^{U,0}\cos\beta\left(1-\cot\psi_u\tan\beta\right)\nonumber\\
  =&\frac{v}{\sqrt{2}}\eta_1^{U,0}\frac{\cos\beta}{\tan\beta}\left(\tan\psi_u-\tan\beta\right)\,,
\end{align}
and 
\begin{align}
  Y'_U=&\frac{v}{\sqrt{2}}\left(-\eta_1^{U,0}\sin\beta+\eta_2^{U,0}\cos\beta\right)\nonumber\\
  =&\frac{v}{\sqrt{2}}\eta_1^{U,0}\cos\beta\left(-\tan\beta-\cot\psi_u\right)\nonumber\\
  =&-\frac{v}{\sqrt{2}}\eta_1^{U,0}\frac{\cos\beta}{\tan\psi_u}\left(\tan\beta\tan\psi_u+1\right)\nonumber \, \, .\\
\end{align}
Therefore, the alignment condition is
\begin{align}
  Y'_U=&-\frac{1+\tan\psi_u\tan\beta}{\tan\psi_u-\tan\beta}M_U'\nonumber\\
  =&-\cot(\psi_u-\beta)M_U'\nonumber\\
  =&\cot(\beta-\psi_u)M_U'\,.
\end{align}
and it allows to establish a relationship between the parameterizations given by
\begin{align}
  \zeta_U=\cot(\beta-\psi_u)\,,
\end{align}
while from Eq.~\eqref{eq:17} follows 
\begin{align}
  \zeta_F=-\tan(\beta-\psi_F)\,,
\end{align}
where ($F=D,L$).
Using the current parameterization which is $Y_{F}=-\tan(\beta-\psi_F)M_F$ 
and $Y_U=\cot(\beta-\psi_u) M_U$, the limit of the THDM-II corresponds to
\begin{align}
  \psi_u=\psi_d=\psi_e=0\,.
\end{align}

Accordingly, it is possible to establish a relationship between the Cheng and Sher \cite{Cheng:1987rs} parameterization 
and the new proposed
alignment notation using  Eqs.~\eqref{eq:9} and ~\eqref{chengsher} and assuming that at high 
energy cut--off scale the two Yukawa couplings for each fermion type are aligned. Then, they are simultaneously 
diagonalized and FCNCs are absent at three level. Establishing the equalities 
\begin{align}
  \lambda^F_{33}(\Lambda)=\tan(\psi_F-\beta)\,,\,\,\, \lambda^u_{33}(\Lambda)=\cot(\beta-\psi_u)
\end{align}
and $\lambda^f_{ij}(\Lambda)=0$ for $i\neq j$. Non zero non-diagonal couplings can be generated when 
 are evolved from the scale $\Lambda$ to
 the electroweak scale  \cite{Braeuninger:2010td}.

\begin{table}
  \centering
  \begin{tabular}{|l|l|l|l|}
\hline
    &New &  Pich {\it et.al} & Ibarra {\it et.al.}\\\hline
   $X_u$ &$-\cot\psi_u$ &$\xi_U^*e^{i\theta}$ &$\tan\xi_U$\\
   $X_b$ &$\tan\xi_b$ &$\xi_De^{-i\theta}$ &$\tan\xi_D$\\
   $X_e$ &$\tan\xi_e$ &$\xi_Ee^{-i\theta}$ &$\tan\xi_E$\\
\hline  
\end{tabular}
  \caption{Alignment conditions defined as $\eta^{f,0}_2=X_f\eta^{f,0}_1$}
  \label{tab:1}
\end{table}

\begin{table}
  \centering
  \begin{tabular}{|l|l|l|l|l|l|}
\hline
 Alignment &\small{New }&  \small{by \cite{Pich:2009sp,Jung:2010ik}} &\small{by \cite{Braeuninger:2010td}} &\small{by \cite{Davidson:2005cw}} & \small{by \cite{Cheng:1987rs}}\\\hline
  $X_u$&$\cot(\beta-\psi_u)$  &$\zeta_U^*$ &$-\tan(\beta-\psi_u)$&$\cot\beta_U$&$\lambda_{tt}$\\ 
  $X_d$&$-\tan(\beta-\psi_d)$ &$\zeta_D$ &$-\tan(\beta-\psi_d)$&$-\tan\beta_U$&$\lambda_{bb}$\\ 
  $X_e$&$-\tan(\beta-\psi_e)$ &$\zeta_E$ &$-\tan(\beta-\psi_e)$&$-\tan\beta_E$&$\lambda_{\tau\tau}$\\ 
\hline
  \end{tabular}
  \caption{$\tan\beta$--like parameters defined as $Y_f=X_f M_f$. The columns are obtained 
from the alignment conditions in Table \ref{tab:1}}
\label{tab:2}
\end{table}

As a summary, we present in Table \ref{tab:1} three different parameterizations for the alignment hypothesis. Table \ref{tab:2} contains the relationships between the meaningful parameters on the models already presented by
Haber and Davidson \cite{Davidson:2005cw}, Cheng and Sher \cite{Cheng:1987rs}, Pich {\it et. al.} \cite{Pich:2009sp,Jung:2010ik}, Ibarra
 {\it et. al.} \cite{Braeuninger:2010td} plus our modification.

In the second part of this work we focus on how to differentiate between the THDM type-II and the THDM type III, where we will  use the parameterizations shown in Table \ref{tab:2}. One way to
do this is to compare the predictions for phenomenological processes between the models. We are going to use processes where the charged Higgs boson is involved. The usual processes to look for
the charged Higgs boson experimentally  are the decays $H^+ \to t \bar b$ and $H^+ \to \tau^+ \bar \nu_\tau$. We propose the following ratio 
between these typical processes and compare them in the frameworks of THDM type-II and type III: 
\begin{equation}
\label{ratio}
 R_ {H^+}=\frac{B(H^+ \to \tau^+ \bar \nu_\tau)}{B(H^+ \to t \bar b)}=\frac{\Gamma(H^+ \to \tau^+ \bar \nu_\tau)}{\Gamma(H^+ \to t \bar b)} \, ,
\end{equation}
and we evaluate it in the framework of the THDM type-III. The decay widths are explicitly

\begin{eqnarray}
\Gamma \left( H^{+}\rightarrow t\bar{b}\right) _{III} &=& \frac{%
3m_{H^{+}}K_{tb}^{2}}{16\pi v^{2}}\left[ 
\left(1-\frac{m_{t}^{2}}{m_{H^{+}}^{2}}\right) \left( \cot(\beta-\psi_u)^{2}m_{t}^{2}+\tan(\beta-\psi_d)^{2}m_{b}^{2}\right) \right.
\nonumber \\
&& \left. -4\cot(\beta-\psi_u)\tan(\beta-\psi_d)\frac{m_{t}^{2}m_{b}^{2}}{m_{H^{+}}^{2}}
\right]\left(1-\frac{m_{t}^{2}}{m_{H^{+}}^{2}}\right)  \nonumber \\
\Gamma \left( H^{+}\longrightarrow \tau \nu _{\tau }\right) _{III} &=& \frac{%
m_{H}^{+} m_\tau^2}{16\pi v^2 }\left( 1-\frac{m_{\tau }^{2}}{m_{H^{+}}^{2}}\right)
^{2}\tan(\beta-\psi_e)^{2} \,\,\, .
\end{eqnarray}

Using Table \ref{tab:2} it is straightforward to write down these equations in the different parameterizations of the THDM type type-III. Notice that the $R_{H^+}$ expression will be in terms of
the three independent parameters $\tan(\beta-\psi_f)$, it is not important how they have been called.
The space of parameters of the THDM type-III is without doubt bigger than the space of parameters of the THDM type-II where only
one $\tan \beta$ like parameter appears.
 In order to present the predictions for this $R_{H^+}$ in the THDM type-II we evaluate the special case
when $\psi_u=\psi_d=\psi_e=0$, 
using the new parameterization introduced in Table \ref{tab:2}. 

For the purposes of this work, we will assume that the full Higgs spectrum of the THDM has been experimentally established, and that at least one  of the $\tan(\beta)$--like parameter has been measured with some uncertainty. Then, by using the available constraints from flavor physics, we will calculate the expected deviations of the observable $R_{H^+}$ from its THDM type-II value. Instead of exploring the full space of parameters of the THDM type-III model, we will focus on the limiting cases where at least two of the $\tan\beta$ like parameters are equal. In this way,
the THDM type-III under the alignment hypothesis should be explored in such a way that the usual THDM type-II can be clearly identified to allow comparisons.
Therefore, we are going to consider three different THDM-II-like models, these are based on the fact that the parameters
$\tan \beta_d$ or $\tan \beta_e$ will be established experimentally more easier than the others.  This could be done considering considering the following three different cases:
\begin{itemize}
\item  $\tan(\beta-\psi_d)=\tan(\beta-\psi_e)$. Here $\tan\beta=\tan(\beta-\psi_d)$ will be used as input in the physical basis~\cite{Eriksson:2009ws} of one specific point of the THDM type-II. Switching to the Higgs basis~\cite{Davidson:2005cw}, we will determine the range of $\tan(\beta-\psi_u)$ allowed from the constraints in flavor observables and from there we will quantify the expected deviations of $R_{H^+}$ from the THDM type-II expectations. The procedure will be repeated for each set of the three $\tan\beta$--like parameters allowed from the constraints in flavor observables, to be specified below.
\item $\cot(\beta-\psi_u)=-\tan(\beta-\psi_d)$. We will apply the previous procedure but with  $\tan(\beta-\psi_e)$ instead of $\tan(\beta-\psi_u)$.
\item $\cot(\beta-\psi_u)=-\tan(\beta-\psi_e)$, the starting point to define the THDM-type-II is in this case $\tan\beta=\tan(\beta-\psi_e)$.
\end{itemize}
Notice that Higgs sector of the MSSM corresponds to the third case where the deviations from THDM-II are in the down sector \cite{Davidson:2005cw,Guasch:2001wv,Assamagan:2004wq}.

 Before we deal with the ratio in equation ~\eqref{ratio}, it is important to look for the available space of parameters already constrained using the low energy phenomenology.
 Constraints  have been
already evaluated using flavor-changing processes at low energies in the THDM model with Yukawa Alignment~\cite{Jung:2010ik,Mahmoudi:2009zx} and in more general MFV constructions, where the higher-order powers
of the yukawas are included MFV THDM models~\cite{Buras:2010mh}. 

In order to take into account many of these observables we have used the available software  2HDM-Calculator \cite{Eriksson:2009ws} and  SuperIso
\cite{Mahmoudi:2008tp}. 
As in \cite{Mahmoudi:2009zx} we use the implementation of the Yukawa couplings in the code 2HDM-Calculator ~\cite{Eriksson:2009ws} with
\begin{align}
   [\rho^U]_{ii}=&\cot(\beta-\psi_u)[\kappa^U]_{ii}\,, &[\rho^F]_{ii}=&\tan(\beta-\psi_F)[\kappa^F]_{ii}\qquad F=D,E\,,
\end{align}
and the numerical evaluation of the flavor physics observables with SuperIso v3.2. 
We refer the reader to the original paper \cite{Eriksson:2009ws} for details about the flavor physics constraints used and the set of the input parameters.

\begin{figure}[htp]
\begin{center}
 \includegraphics[scale=0.3]{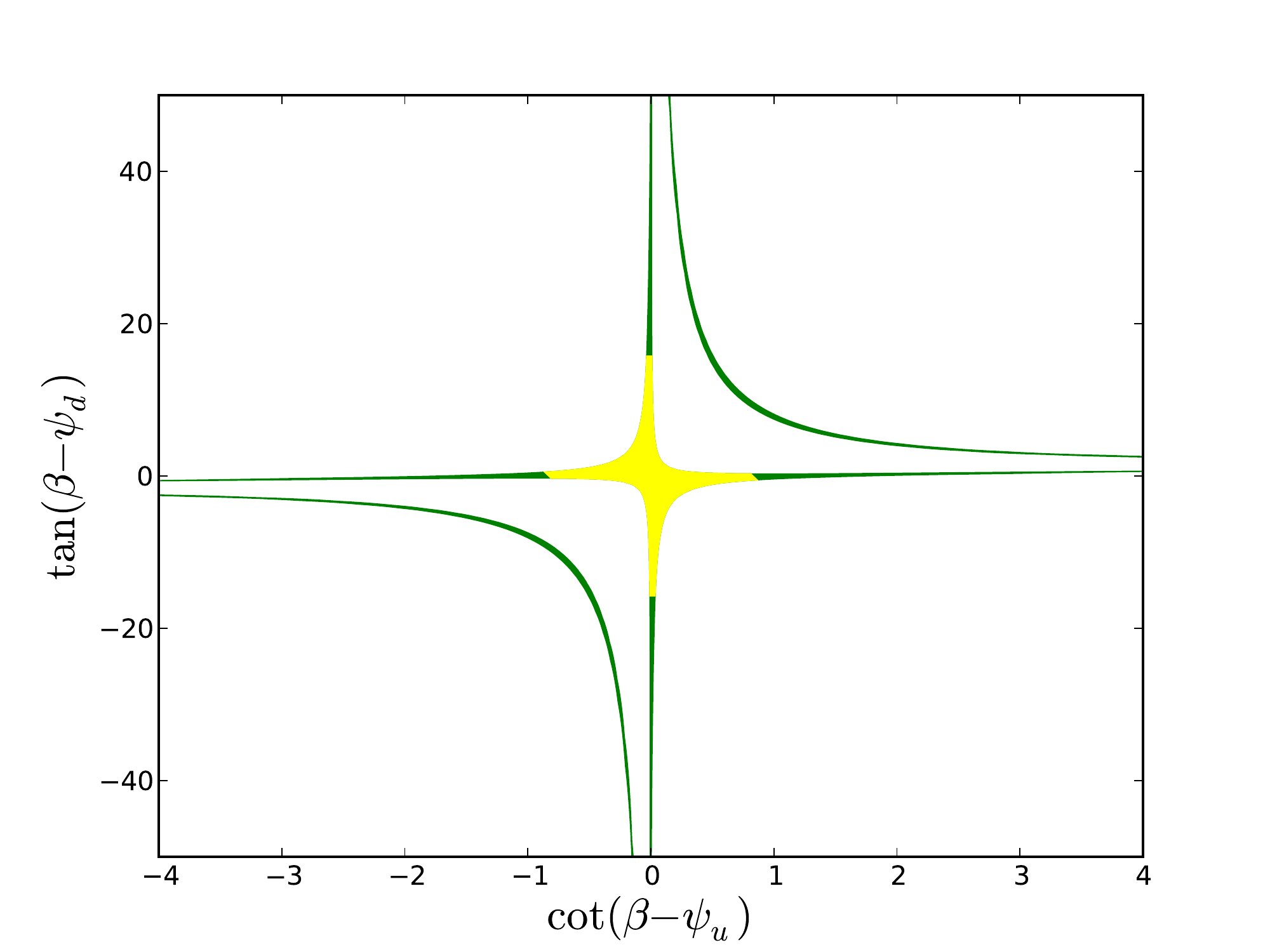} \quad
\includegraphics[scale=0.3]{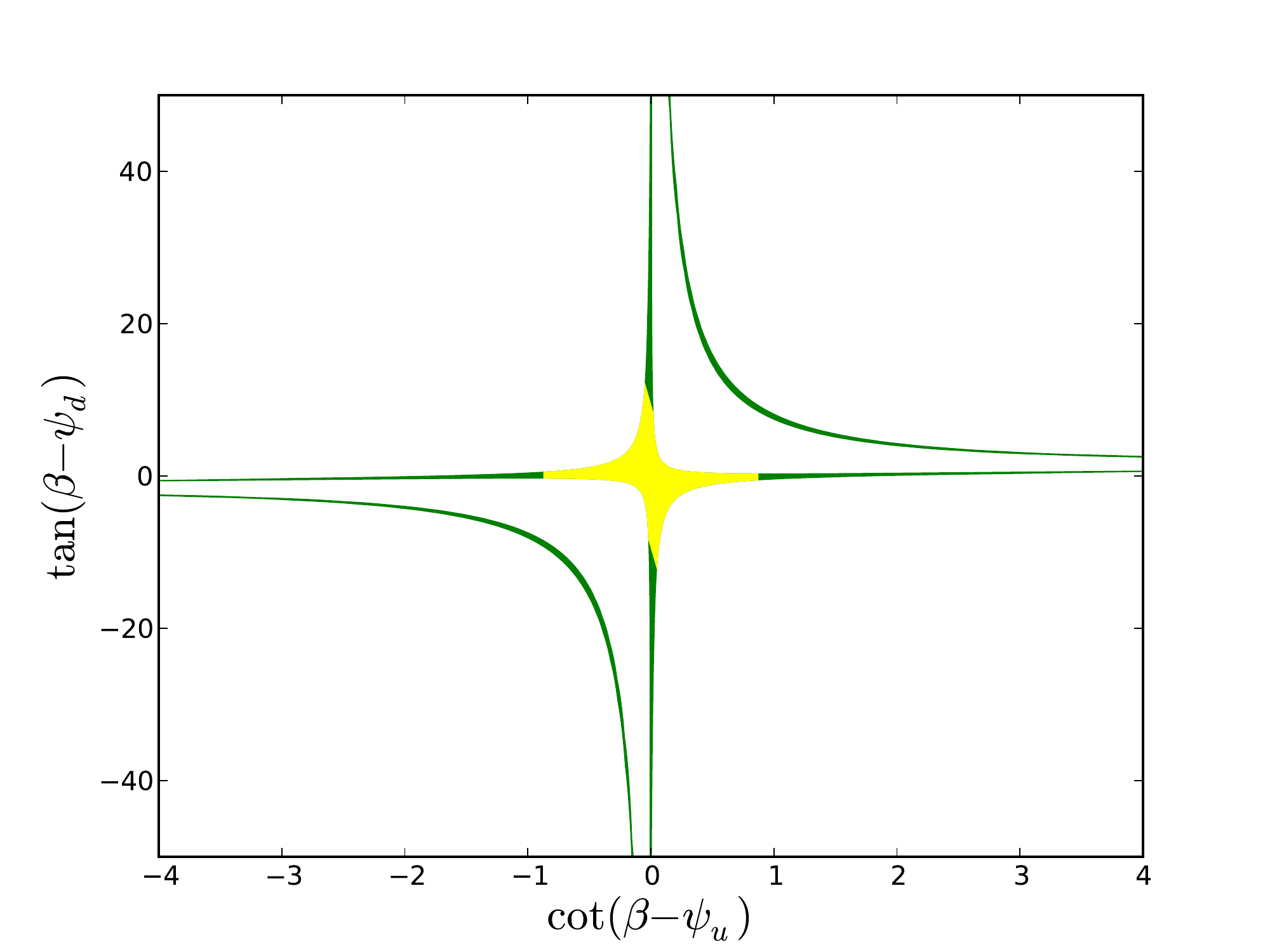} \quad
	 \includegraphics[scale=0.3]{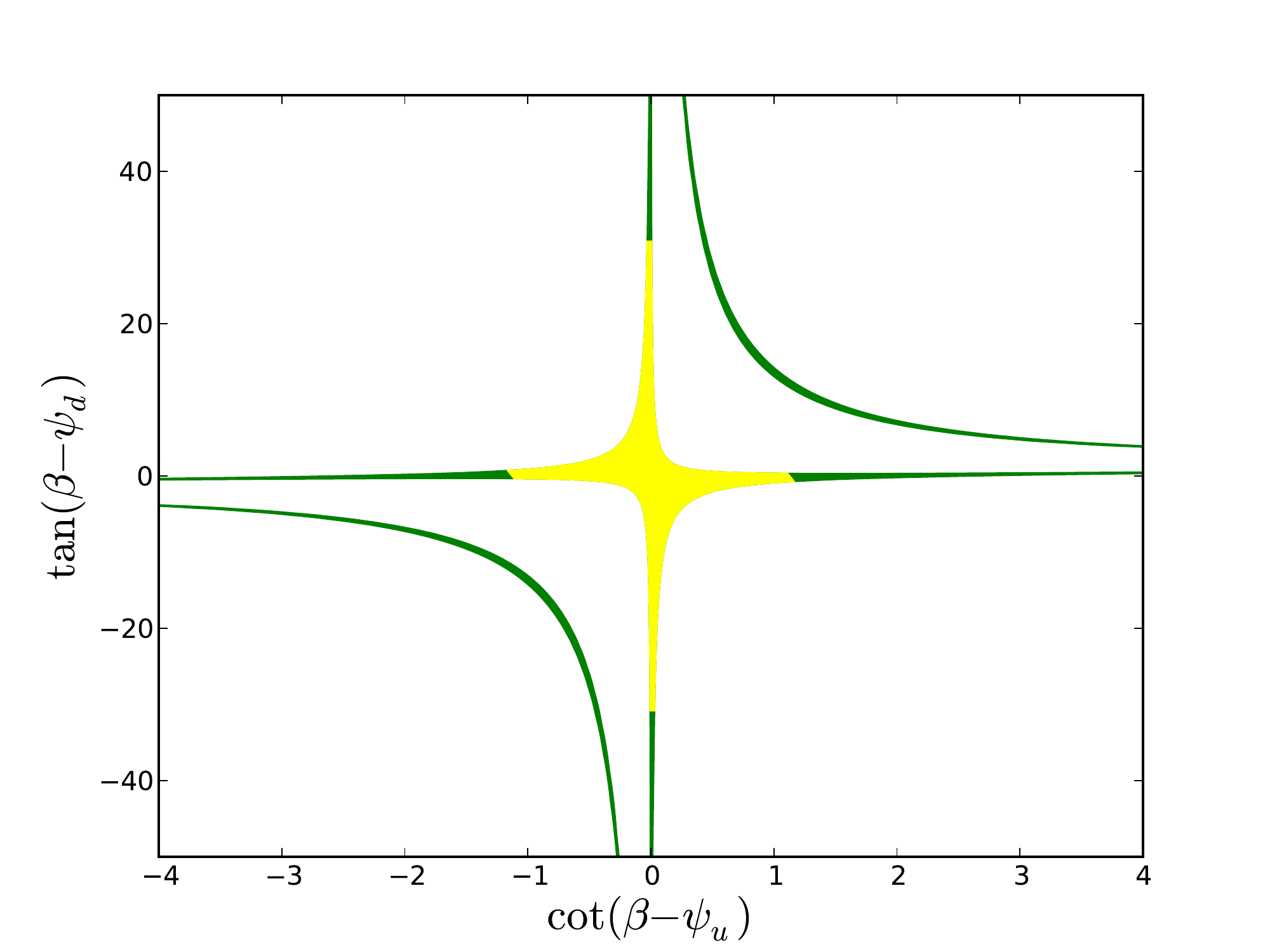} \quad
 \includegraphics[scale=0.3]{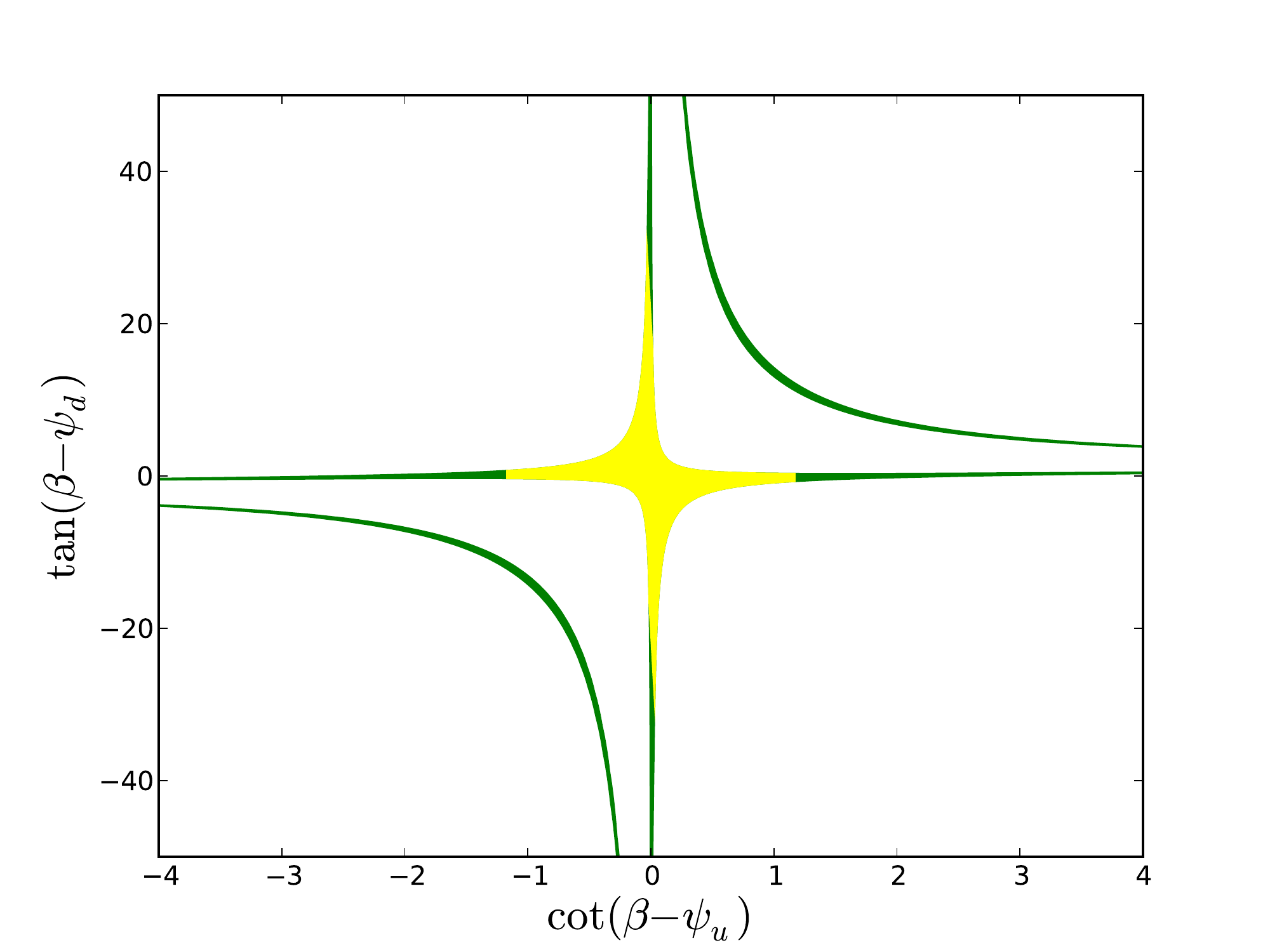} \quad 
\includegraphics[scale=0.3]{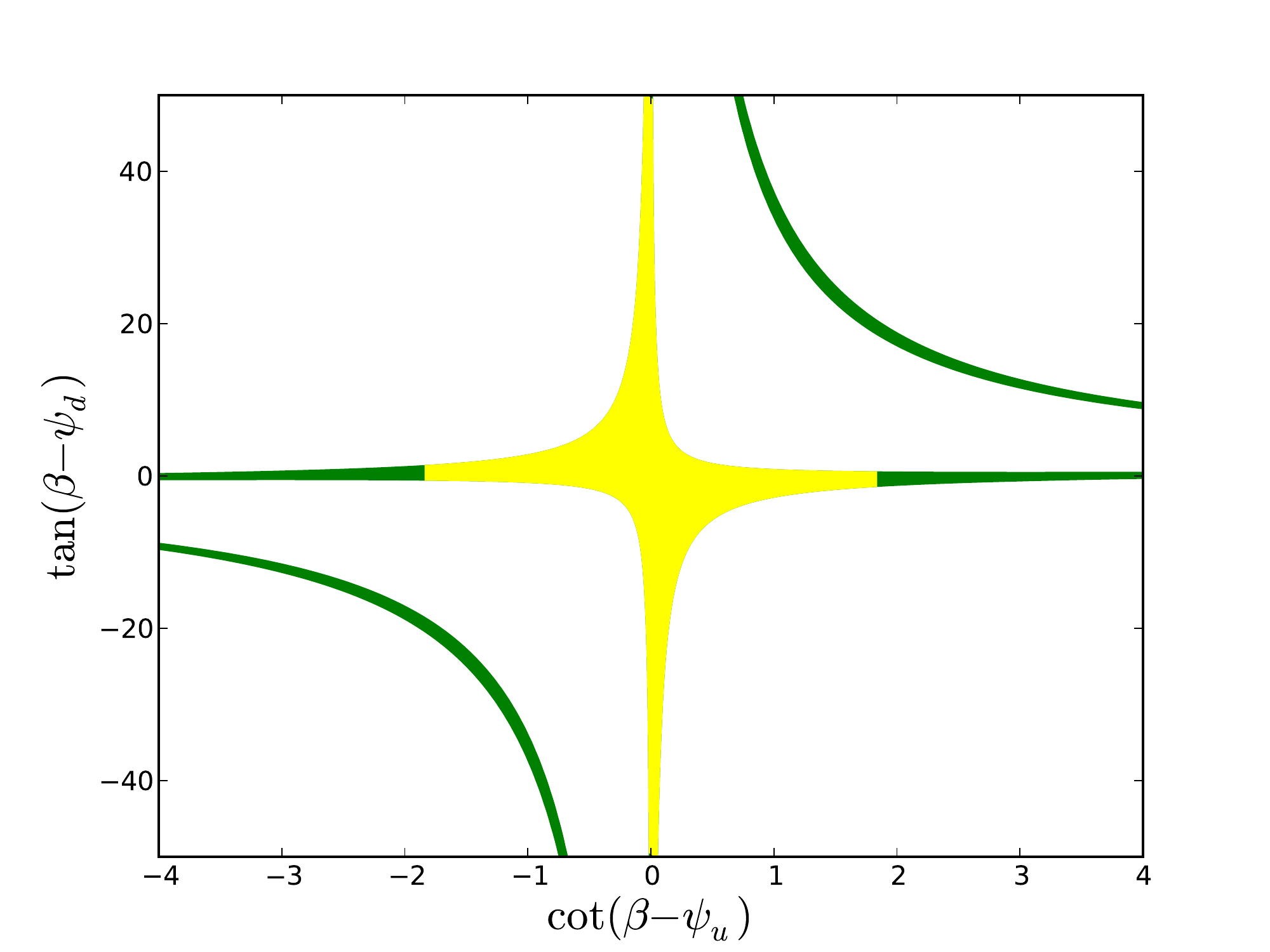} \quad
\includegraphics[scale=0.3]{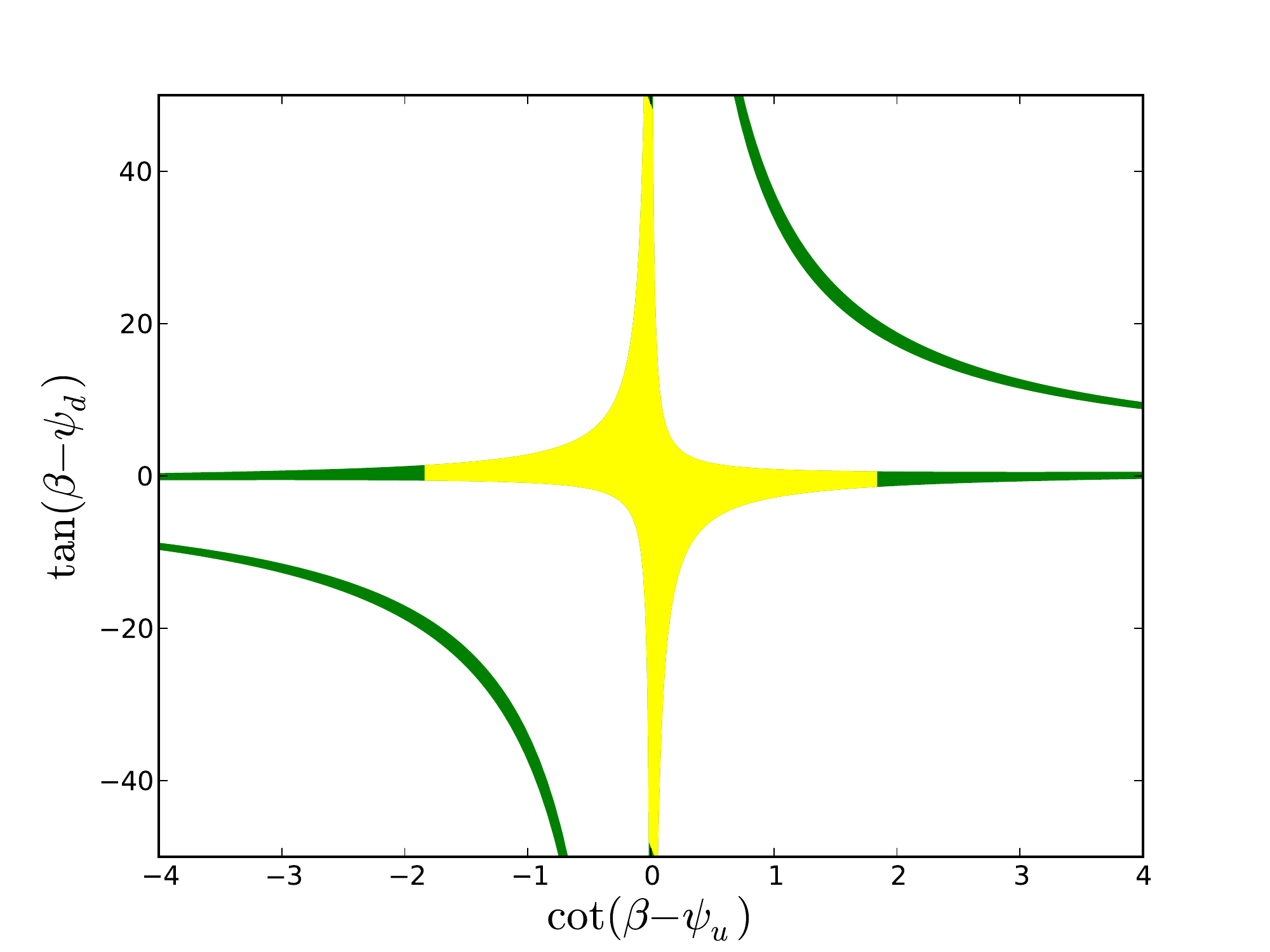}

	    \end{center}
\caption{Constraints from $B \to X_s \gamma$ and the decay $B_s \to \mu^+ \mu^-$.
 The allowed region is the yellow one. Two different cases are presented: $\tan(\beta-\psi_d)=\tan(\beta-\psi_e)$ (left side) and 
$\cot(\beta-\psi_u)=-\tan(\beta-\psi_e)$} (right side) for the Charged Higgs boson masses 180, 350 and 800 GeV respectively from up to down.
\label{bsgplots}
\end{figure}

\begin{figure}[htp]
\begin{center}
	 \includegraphics[scale=0.3]{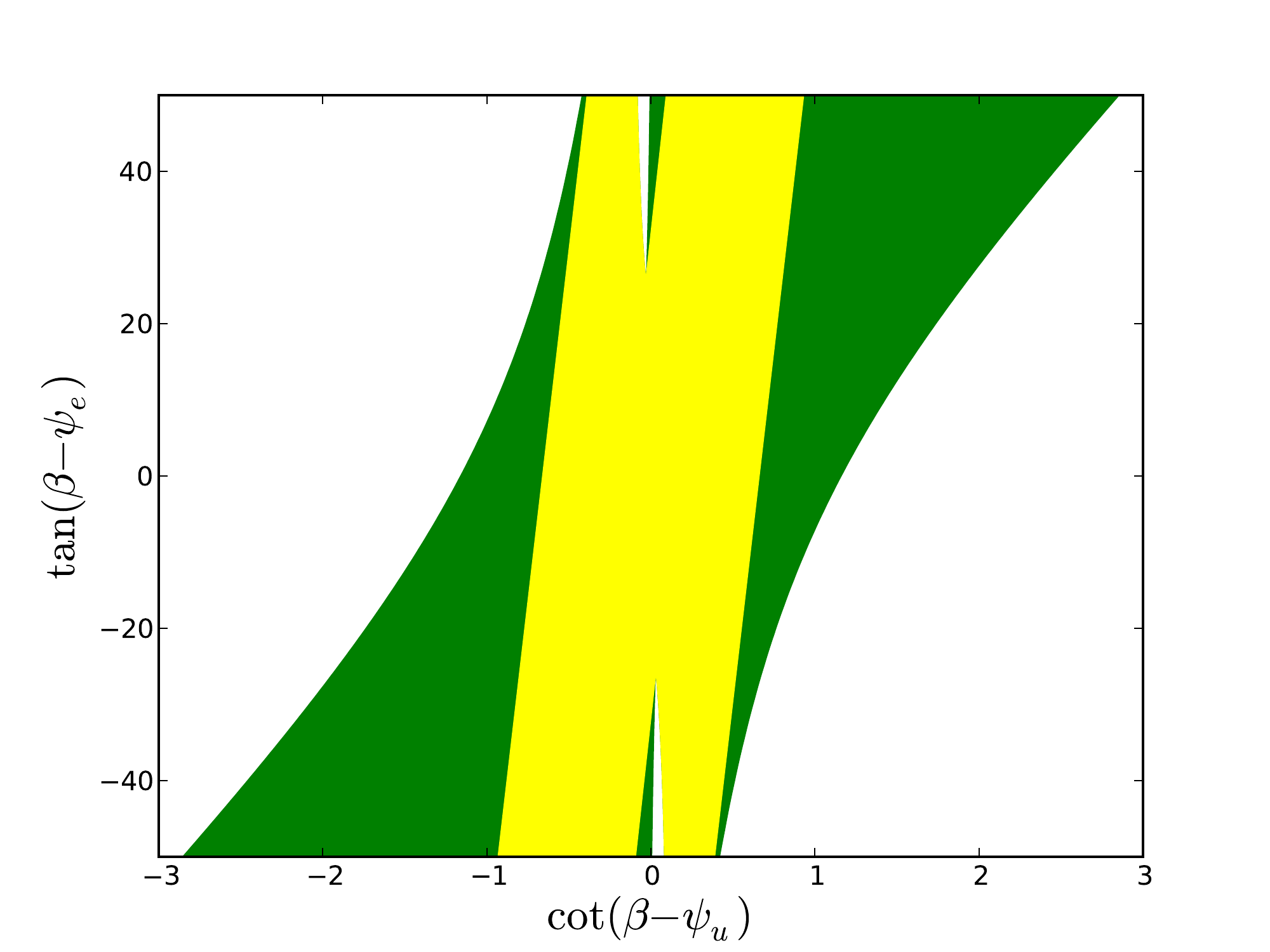} \quad
 \includegraphics[scale=0.3]{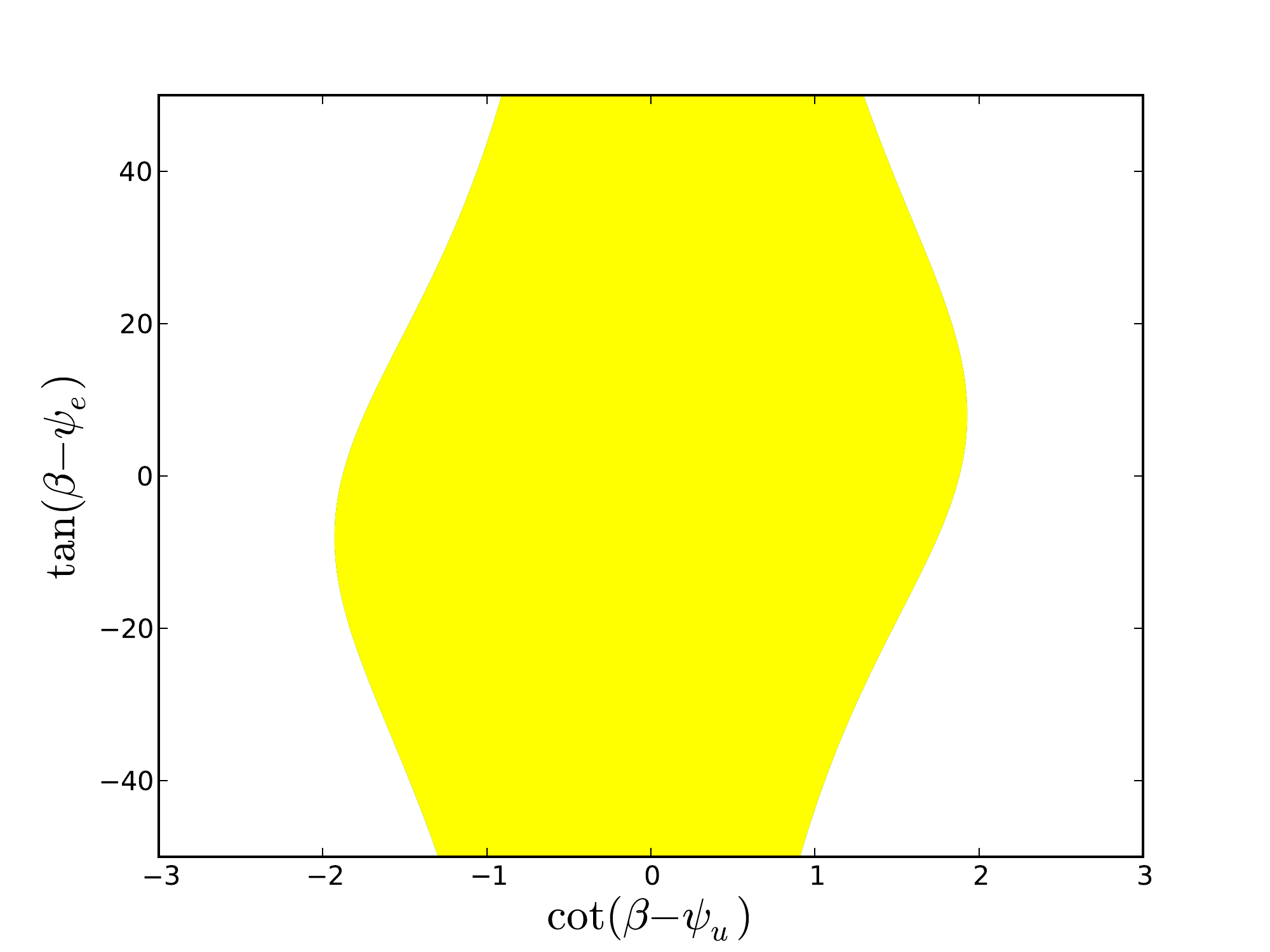} 
	    \end{center}
\caption{Constraints from $B_s \to \mu^+ \mu^-$ and the decay $B_s \to \mu^+ \mu^-$ in the plane $\tan(\beta-\psi_e)-\cot(\beta-\psi_u)$ taking the Charged Higgs mass 350 and 800 GeV.}
\label{bsmumuplots}
\end{figure}

As it has been pointed out in \cite{Pich:2009sp,Jung:2010ik,Mahmoudi:2009zx},   the most restrictive processes in the ATHDM are $B \to X_s \gamma$, $\Delta_0(B\to K^*\gamma)$, $B_u\to \tau\nu_\tau$, and $Z\to \bar{b}b$. In this work we also take into account the strong constraints on $B \to \mu^+ \mu^-$ reported recently by the  LHCb collaboration \cite{Aaij:2012ac} which help to constraint even more the space of parameters. The results are presented in figure \ref{bsgplots}. 
Two different cases are presented: $\tan(\beta-\psi_d)=\tan(\beta-\psi_e)$ (left side) and
$\cot(\beta-\psi_u)=-\tan(\beta-\psi_e)$ (right side) for  the Charged Higgs boson masses 
180, 350 and 800 GeV respectively from up to down.
The allowed regions from $B \to X_s \gamma$ of space of parameters $(\tan(\beta-\psi_e)-\cot(\beta-\psi_u))$ are shown in dark gray (green) in  figure \ref{bsmumuplots} and correspond to those show in references \cite{Pich:2009sp,Jung:2010ik,Mahmoudi:2009zx}. In this work we choose to display the other important constraints in the same $(\tan(\beta-\psi_e)-\cot(\beta-\psi_u))$ plane in order to illustrate their effectiveness. The final allowed region is shown in the light gray (yellow) regions.  The branches with simultaneously large and equal signs values of  $\tan(\beta-\psi_e)$ and $\cot(\beta-\psi_u)$ are excluded~\cite{Mahmoudi:2009zx} from the isospin asymmetry in the
exclusive decay mode $B\to K^*\gamma$ 
\begin{align}
  \Delta_0(B\to K^*\gamma)\equiv
  \frac{\Gamma(\bar{B}^0\to \bar{K}^{*0}\gamma)-\Gamma(\bar{B}^-\to \bar{K}^{*-}\gamma)}{
  \Gamma(\bar{B}^0\to \bar{K}^{*0}\gamma)+\Gamma(\bar{B}^-\to \bar{K}^{*-}\gamma)}\,.
\end{align}
In the horizontal axis, the constraint in the central region (the interval $\sim [-1,1]$)
 is coming 
from $B_s \to \mu^+ \mu^-$ with few changes for the illustrated cases, while in the vertical axis there are strong bounds coming from $B_u\to \tau\nu_\tau$ in the central region.  As a complement
in figure \ref{bsmumuplots}  the plane $\tan(\beta-\psi_e)-\cot(\beta-\psi_u)$ is shown taking the
 Charged Higgs boson masses 350 and 800 GeV, the green (dark) region is allowed by $B_s \to \mu^+ \mu^-$ but again processes like $B \to X_s \gamma$ constrain the allowed region only to the central zone. In the calculations for the process $B_s\to \mu^+ \mu^-$,
it is necessary to fix the Higgs mass values in the neutral sector, we have chosen  $m_h=125$ GeV, $m_H=m_{H^+}+10$ GeV and $m_A=m_{H^+}+15$ GeV and
we have found that the constraints coming from  $B_s\to \mu^+ \mu^-$ have a mild dependence on the neutral Higgs masses.

\begin{figure}[hbp]
\begin{center}
 \includegraphics[scale=0.25]{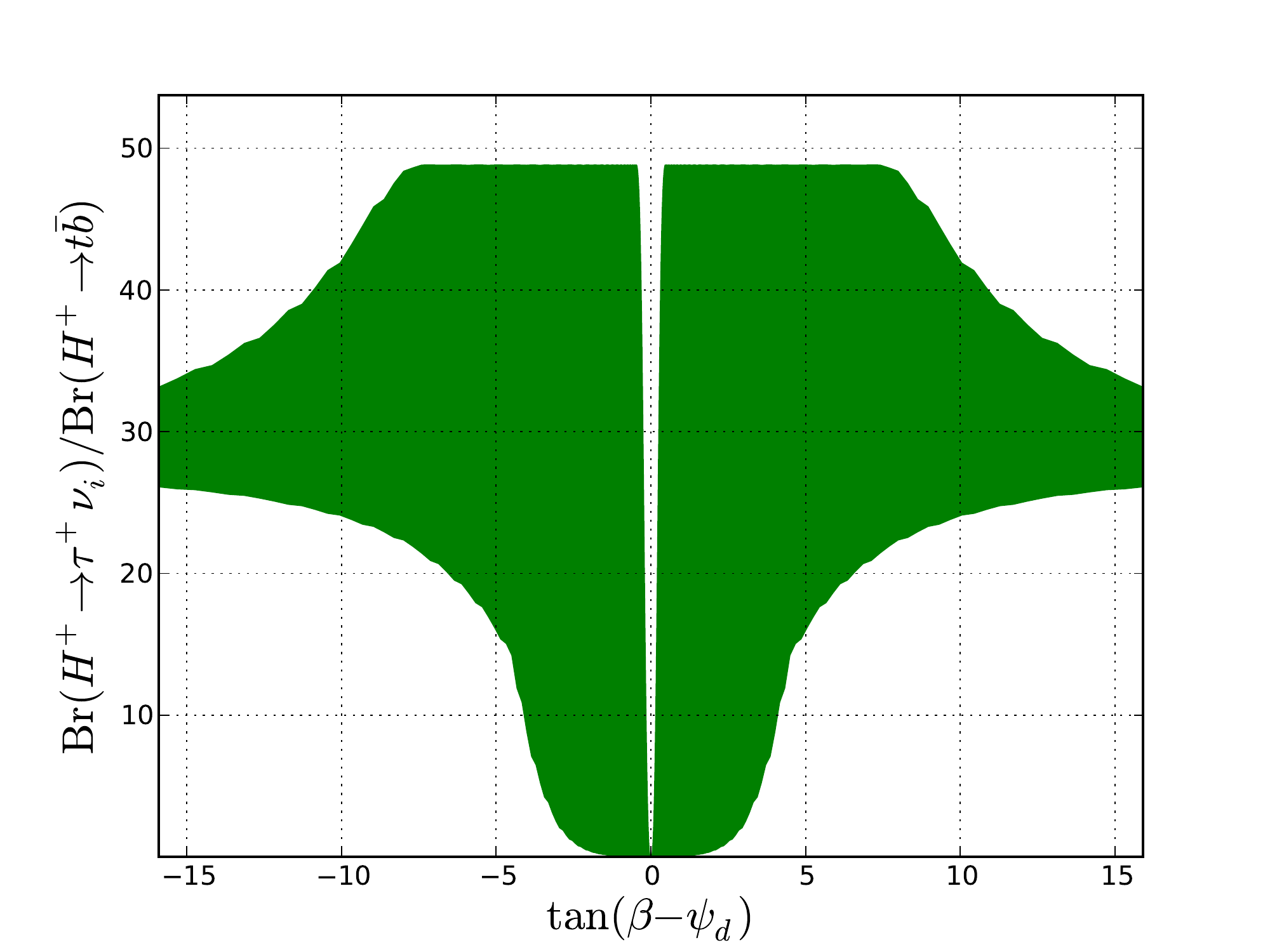} \quad
	 \includegraphics[scale=0.25]{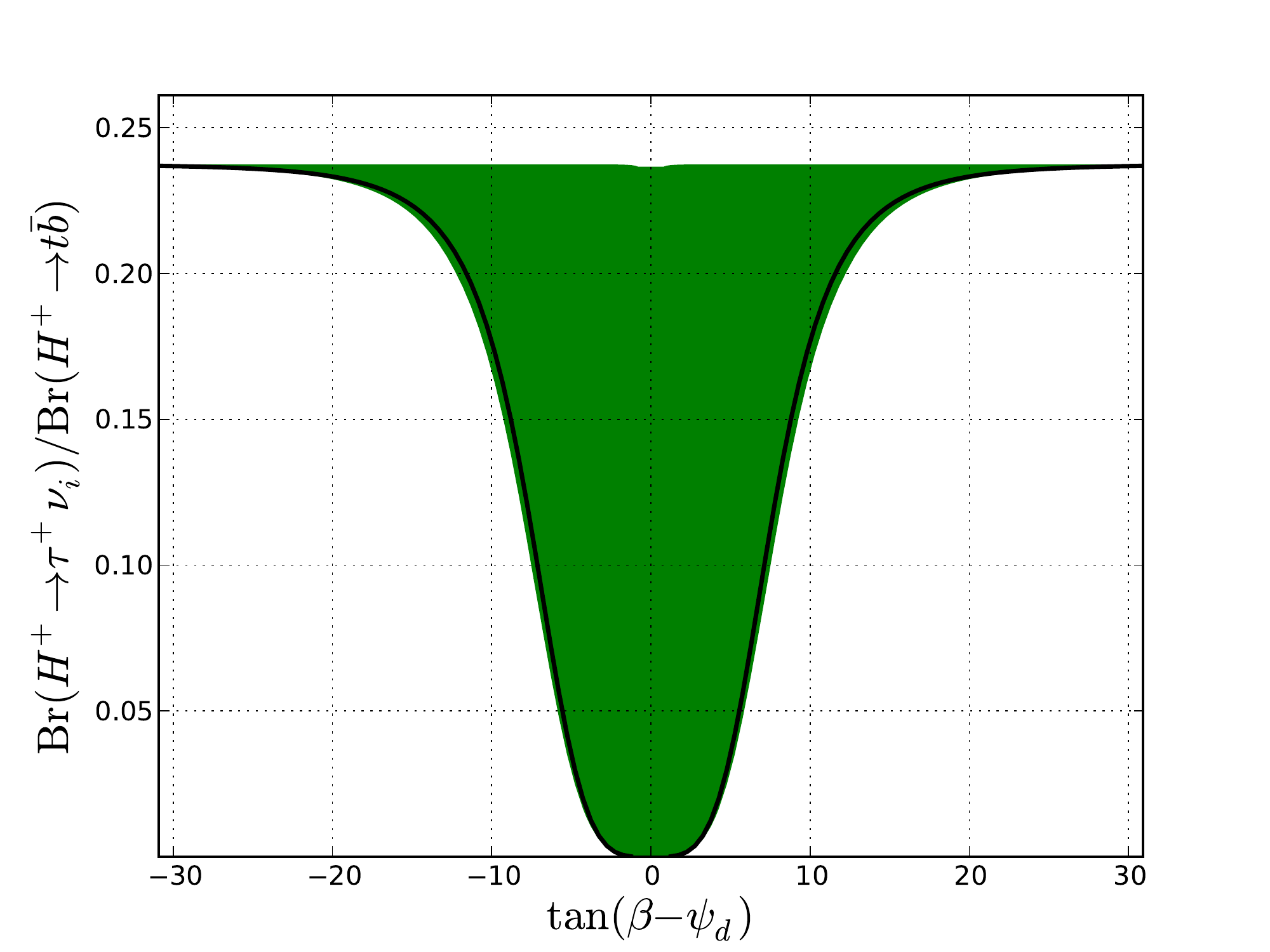} \quad
 \includegraphics[scale=0.25]{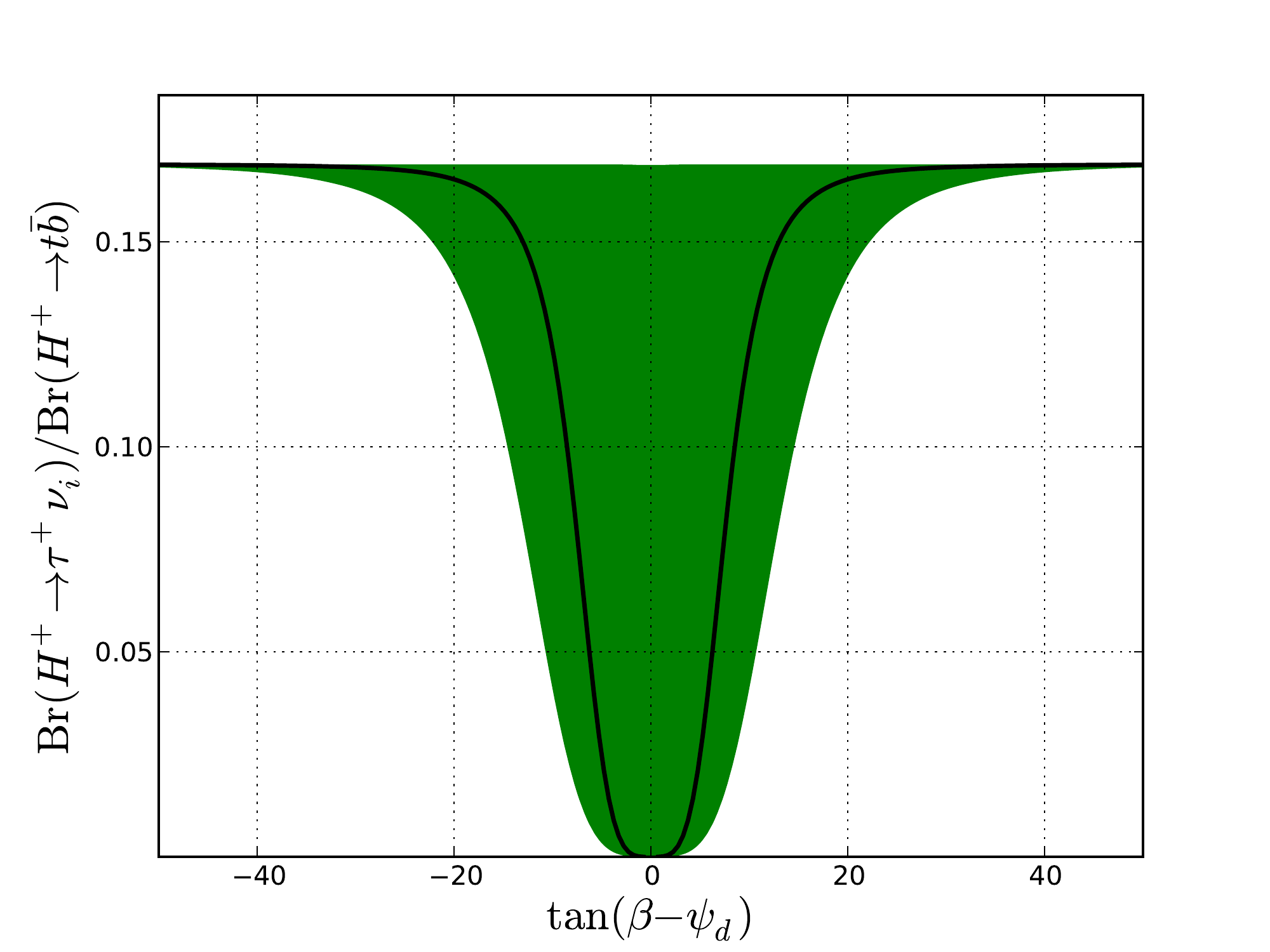} 
	    \end{center}
\caption{the ratio $R_{H^+}$ as a function of the parameter $\tan(\beta-\psi_d)$ for different values
of the charged Higgs boson mass $m_{H^+}=180,350,800$ GeV using the assumption $\tan(\beta-\psi_d)=\tan(\beta-\psi_e)$.}
\label{rh1}
\end{figure}

\begin{figure}[hbp]
\begin{center}
\includegraphics[scale=0.25]{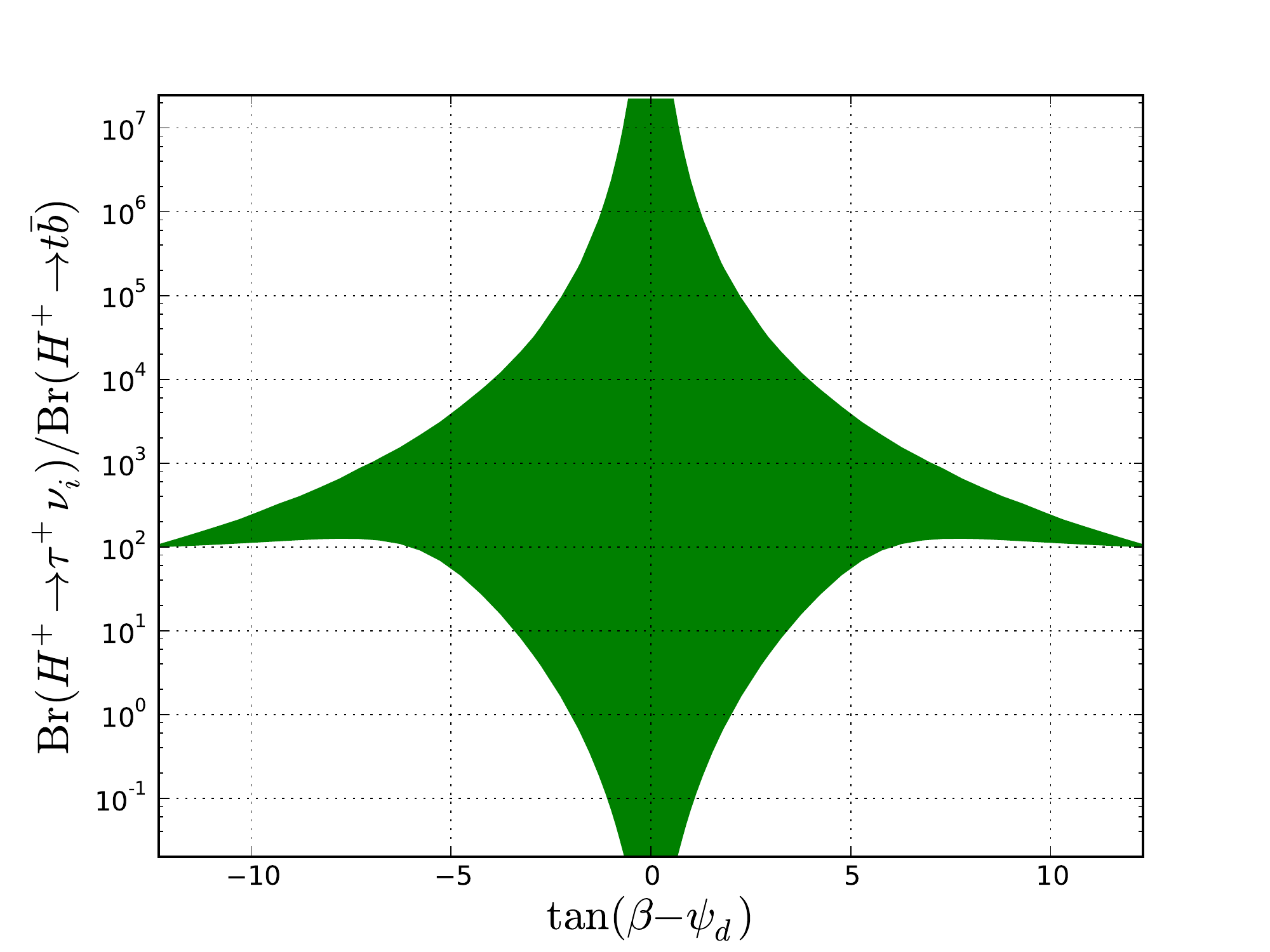} \quad
	 \includegraphics[scale=0.25]{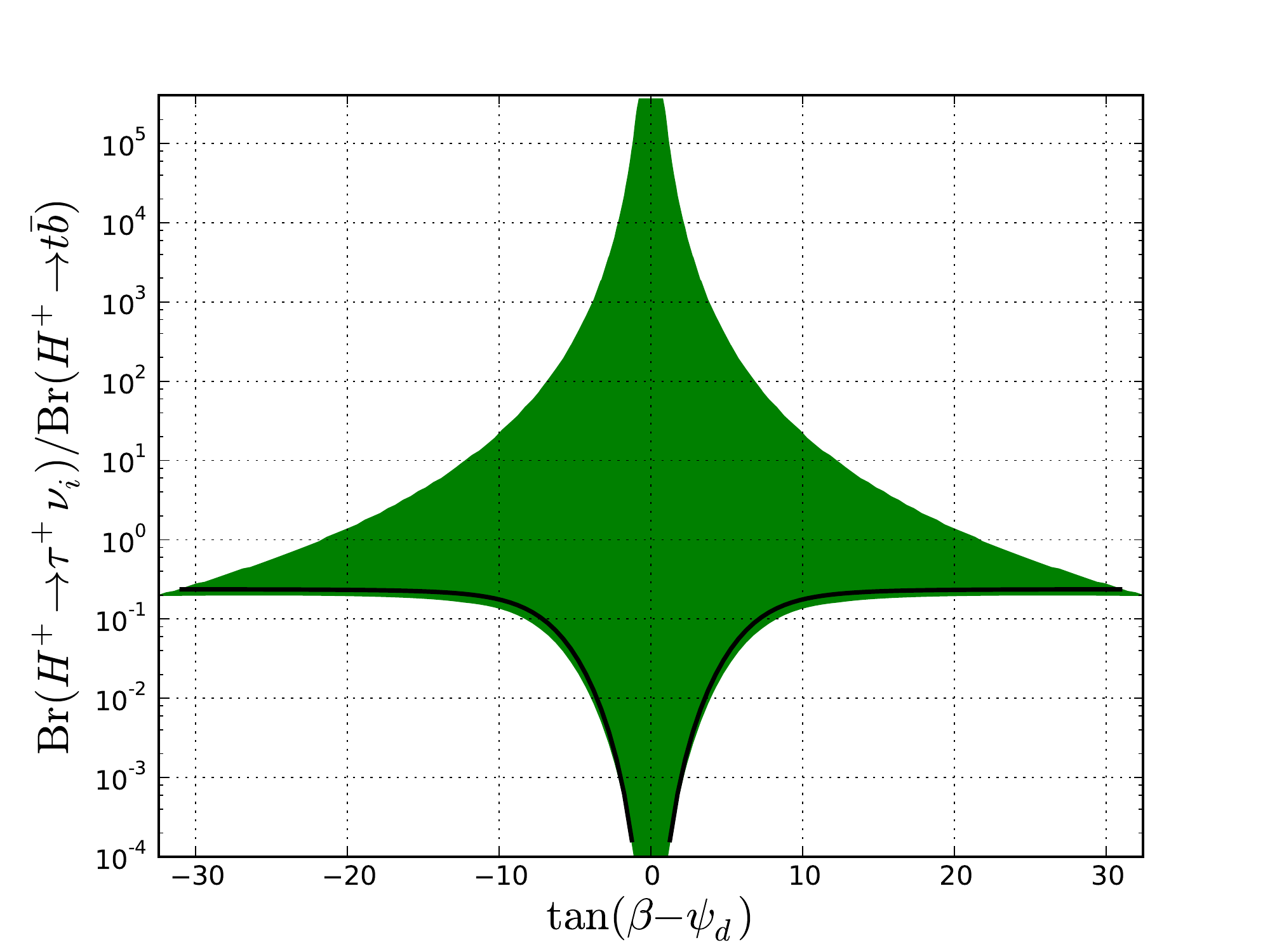} \quad
 \includegraphics[scale=0.25]{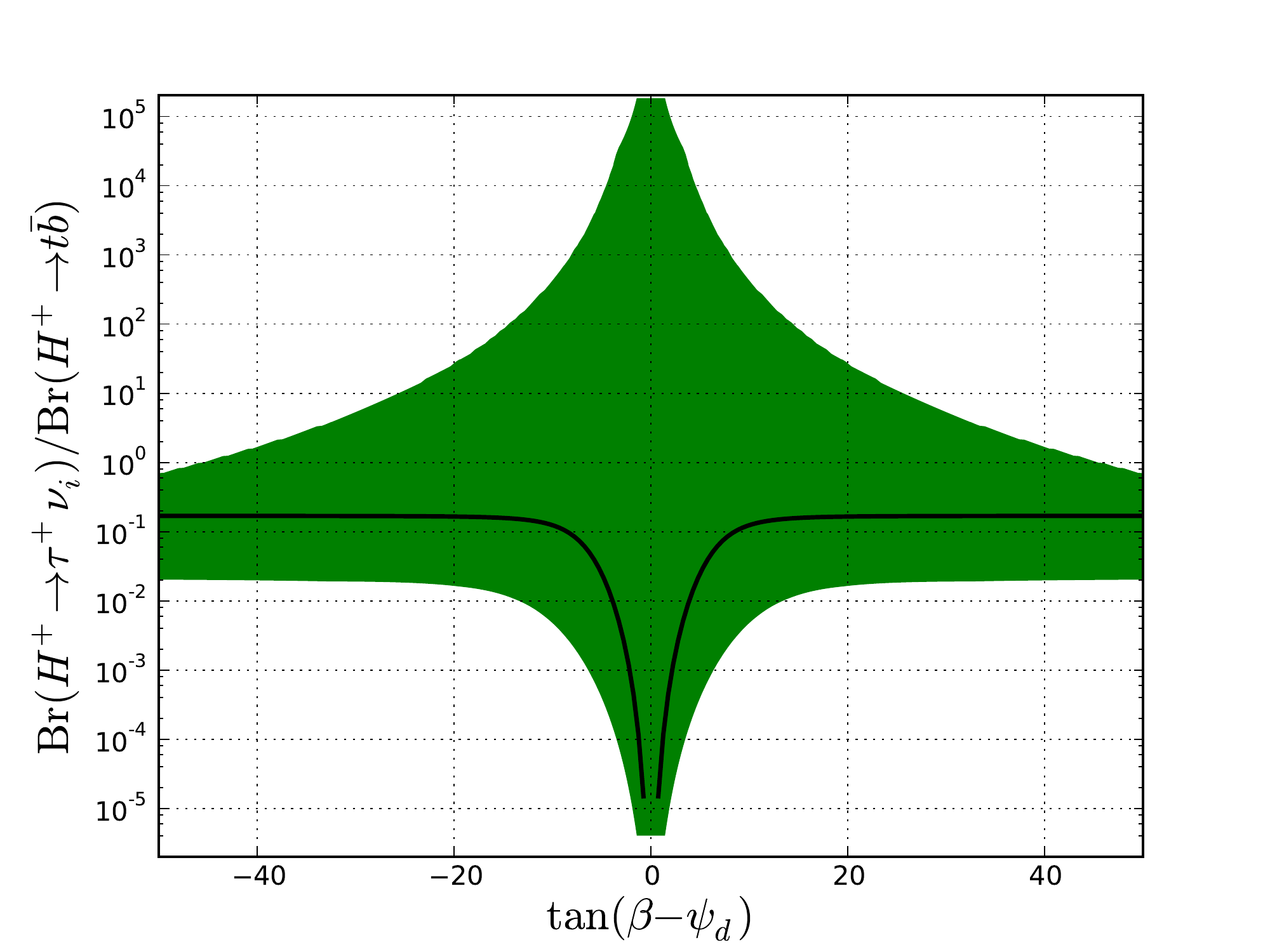} 	
	    \end{center}
\caption{the ratio $R_{H^+}$ as a function of the parameter $\tan(\beta-\psi_d)$ for different values
of the charged Higgs boson mass $m_{H^+}=180,350,800$ GeV using the assumption $\cot(\beta-\psi_u)=-\tan(\beta-\psi_e)$. }
\label{rh2}
\end{figure}

\begin{figure}[hbp]
\begin{center}
 \includegraphics[scale=0.3]{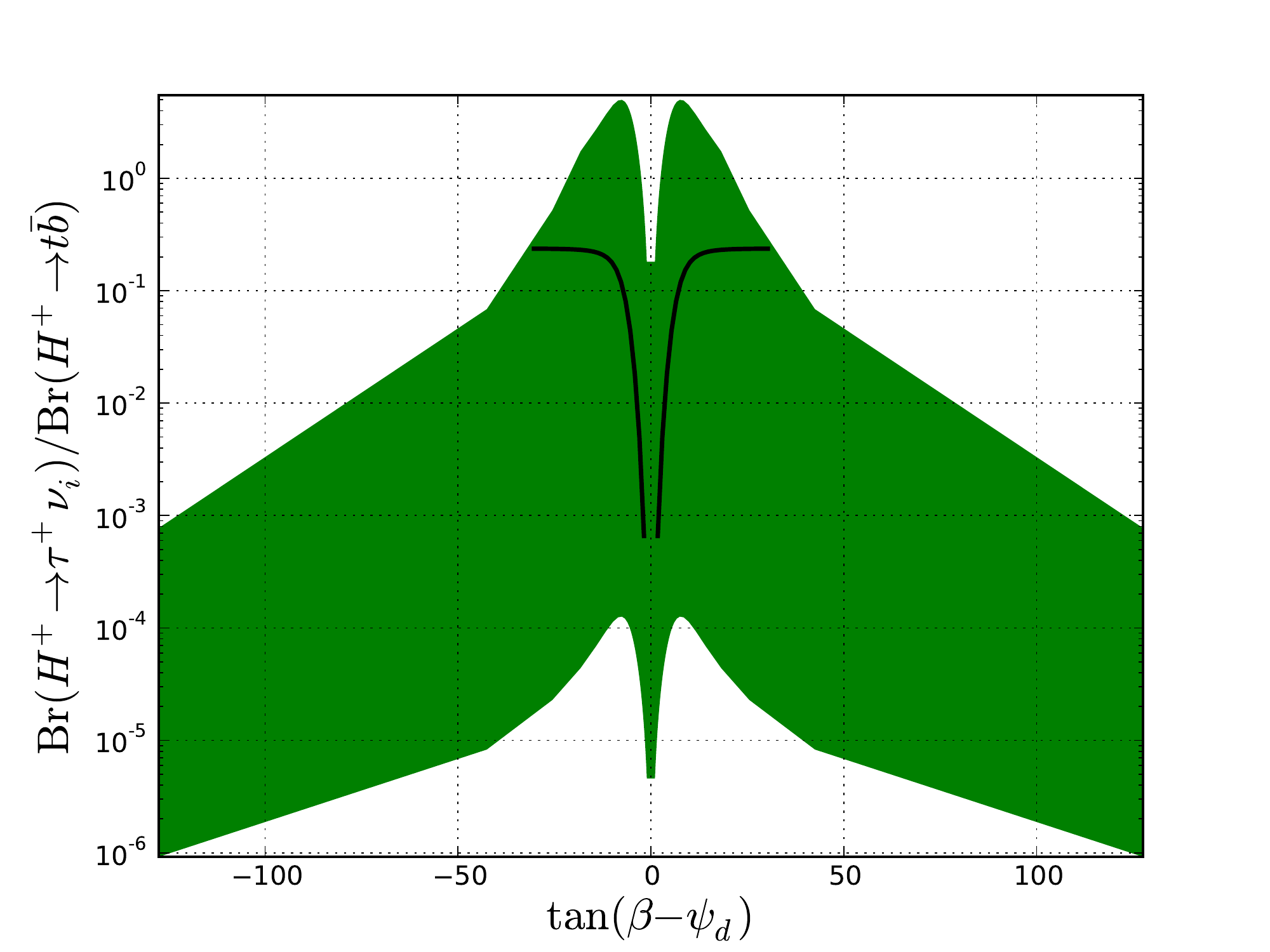} \quad
 \includegraphics[scale=0.3]{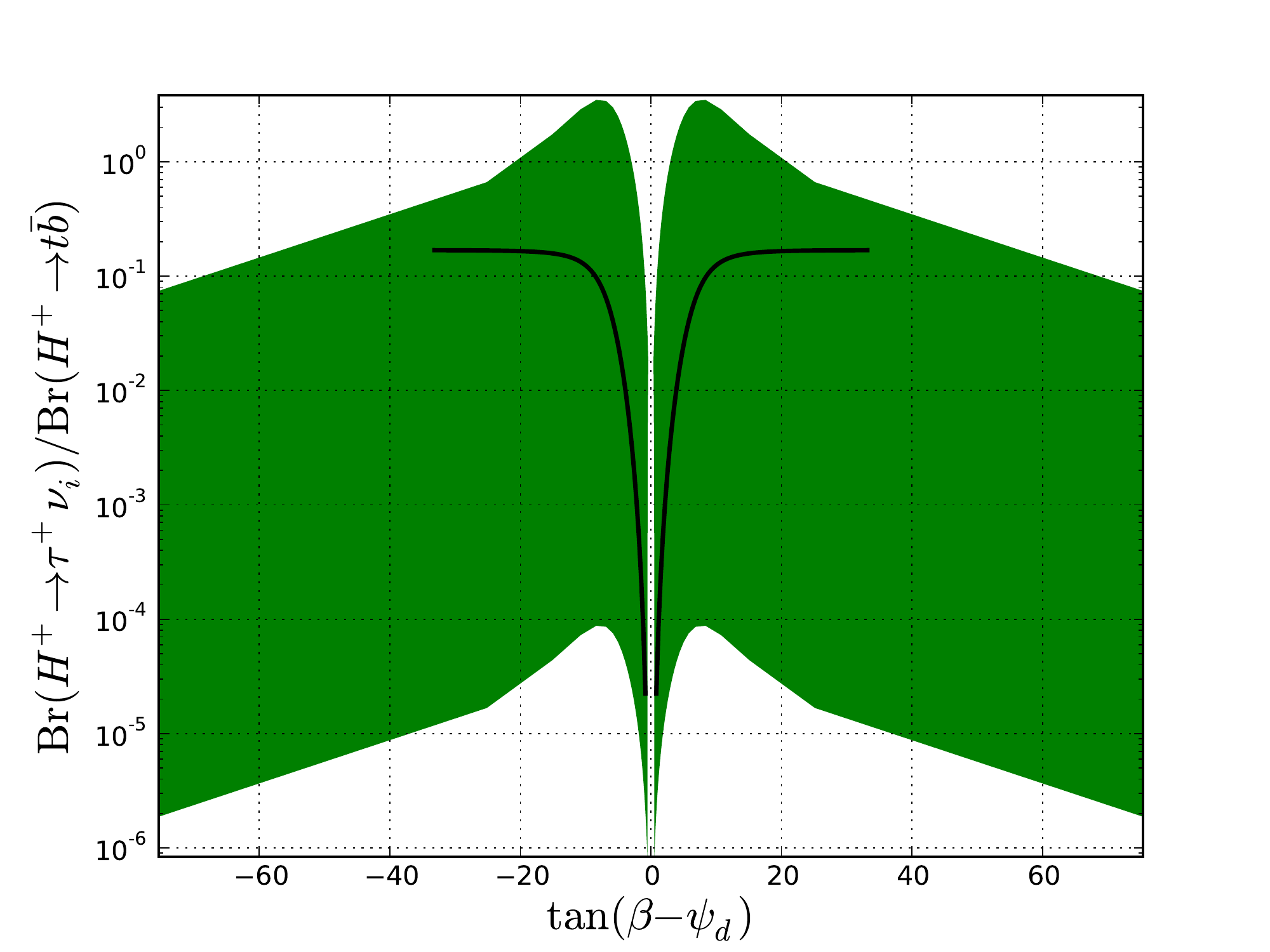} 	
	    \end{center}
\caption{the ratio $R_{H^+}$ as a function of the parameter $\tan(\beta-\psi_d)$ for different values
of the charged Higgs boson mass $m_{H^+}=350,800$ GeV using the assumption $\cot(\beta-\psi_u)=-\tan(\beta-\psi_d)$ and $-0.5 \leq \cot(\beta-\psi_u) \leq 0.5$. }
\label{rh3}
\end{figure}

Taking into account the whole set of constraints, the ratio $R_{H^+}$ ~\eqref{ratio} is plotted as a function of the parameter $\tan(\beta-\psi_d)$ for different values
of the charged Higgs boson mass 180, 350 and 800 GeV. We have considered the three already mentioned cases when $\tan(\beta-\psi_d)=\tan(\beta-\psi_e)$ (figure \ref{rh1}), $-\tan(\beta-\psi_e)=\cot(\beta-\psi_u)$
 (figure \ref{rh2}) and when $\cot(\beta-\psi_u)=-\tan(\beta-\psi_d)$ 
(figure \ref{rh3}). In figures \ref{rh1}-\ref{rh3}, the parameter $\cot(\beta-\psi_u)$ is taking the allowed values obtained
from figures \ref{bsgplots} and \ref{bsmumuplots}.
Also, in these plots, the THDM type-II contribution is shown as a black solid line which is inside the larger allowed region by the THDM type-III.  For a charged Higgs boson mass 
$m_{H^+}=180$ GeV there is not THDM type-II contribution because this mass value is already ruled out in that framework. The observation of light charged Higgs masses or $R_{H^+}>2\times10^{-1}$ are a clear signature of Yukawa structures beyond the THDM type-II. Even for small $R_{H^+}$ an anomalous value of $\tan(\beta-\psi_d)$ with respect to the THDM type-II expectations, could be identified. It is worth to stress that for cases $\cot(\beta-\psi_u)=-\tan(\beta-\psi_e)$ and $\cot(\beta-\psi_u)=-\tan(\beta-\psi_d)$ large deviations from the THDM type-II are expected.

To conclude, different parameterizations of the THDM type-III have been reviewed and compared under the alignment hypothesis. Those parameterizations are
based on the work by 
Haber and Davidson \cite{Davidson:2005cw}, Cheng and Sher \cite{Cheng:1987rs}, Pich {\it et. al.} \cite{Pich:2009sp,Jung:2010ik}, Ibarra
 {\it et. al.} \cite{Braeuninger:2010td} and our modification introduced in Table \ref{tab:1}.  Equivalences between them have been found and they are shown in Table \ref{tab:2}. 
On the other hand, 
different constraints on the space of parameters of the ATHDM have been found. The relevant bounds are coming from 
$B\to X_s \gamma$, $\Delta_0(B\to K^*\gamma)$, $B_u\to \tau\nu_\tau$, and $B_s \to \mu^+ \mu^-$ for the parameters involved in the 
charged Higgs sector. The $B\to X_s \gamma$ process is evaluated at NNLO in this framework and for the process $B_s \to \mu^+ \mu^-$ the most restrictive bound by the LHCb collaboration is used. 
In the general type-III model it is necessary to extract three physical $\tan\beta$--like parameters from collider observables, like the charged Higgs branchings.
In this work we propose the fraction 
 $R_ {H^+}={B(H^+ \to \tau^+ \bar \nu_\tau)}/{B(H^+ \to t \bar b)}$ in order to quantify the differences between them. 
Our results show significant differences between the THDM type II and III, even when a simplified version of the THDM type III is considered (ATHDM). The Lagrangian of these
kind of models (THDM type III and ATHDM) is similar to the effective Lagrangian coming from models with extra symmetries, either discrete symmetries or supersymmetry. By obtaining
the effective Lagrangian, the extra symmetries in those models are usually broken and their $\tan \beta$ parameters cease to be well defined \cite{Davidson:2005cw}. In this work
 we have illustrated the necessity to replace $\tan\beta$ with parameters 
better suited for general phenomenological
and theoretical studies of the scalar sector. If a neutral Higgs is finally 
found at the LHC, the next task could be the determination of the underlying scalar sector.

We acknowledge C. Sandoval for useful discussions and carefully reading of the manuscript.
J.-Alexis R and H. Cardenas have been supported in part by UNAL-DIB-Bogot\'a grant 14844  and D.R has been supported in part by UdeA/2011 grant: IN1614-CE.

\end{document}